\documentclass[aps,rpl,preprint,groupedaddress]{revtex4}
\begin{document}

\title{Entanglement teleportation using three-qubit entanglement}
\author{Ye Yeo}

\affiliation{Centre for Mathematical Sciences, Wilberforce Road, Cambridge CB3 0WB, United Kingdom}

\begin{abstract}
We investigate the teleportation of an entangled two-qubit state using three-qubit entangled GHZ and W channels.  The effects of white noise on the average teleportation fidelity and amount of entanglement transmitted are also studied.
\end{abstract}

\maketitle

The linearity of quantum mechanics allows building of superposition states of composite system that cannot be written as products of states of each subsystem.  Such states are called entangled.  States which are not entangled are referred to as separable states.  An entangled composite system gives rise to nonlocal correlation between its subsystems that does not exist classically.  This nonlocal property enables the uses of local quantum operations and classical communication to transmit information with advantages no classical communication protocol can offer.  The understanding of entanglement is thus at the very heart of quantum information theory \cite{Nielsen}.  In recent years, three-particle entangled states have been investigated by a number of authors \cite{Vidal, Dur, Acin, Andrianov, Rajagopal}.  They have also been shown to have advantages over the two-particle Bell states in their application to dense coding \cite{Wiesner, Hao}, teleportation \cite{Bennett, Karlsson, Yeo1}, and cloning \cite{Buzek, Ekert}.

In Refs. \cite{Vidal, Dur}, Dur {\it et al.} pointed out that the three-particle entangled GHZ state \cite{Greenberger}, while maximally entangled, is not robust in that if one of the three particles is traced out, the remaining two-particle system is not entangled as measured by several criteria.  On the other hand, another three-particle entangled W state \cite{Zeilinger}, which is inequivalent to the GHZ state under stochastic local operations and classical communication, is robust in that it remains entangled even after any one of the three particles is traced out.  More recently, Sen {\it et al.} \cite{Dagomir} showed that $N$-particle entangled W states, for $N > 10$, lead to more ``robust'' (against white noise admixture) violations of local realism, than $N$-particle entangled GHZ states.  The GHZ and W states thus exhibit very different properties when subjected to physical processes like state loss, or white noise.

In Ref.\cite{Gorbachev}, V. N. Gorbachev and A. I. Trubilko demonstrated a teleportation scheme $P_0$ for an entangled two-qubit state to be transmitted using a three-qubit GHZ channel.  L. Marinetto and T. Weber \cite{Weber} considered the same protocol $P_0$ and the problem of which kind of two-qubit states can be perfectly teleported through a three-qubit quantum channel.  B. S. Shi, Y. K. Jiang and G. C. Guo \cite{Shi} developed conditional teleportation protocols for transmitting two entangled qubits using non-maximally entangled three-qubit GHZ channels.  In this paper, we review the Gorbachev-Trubilko entanglement teleportation scheme $P_0$ \cite{Gorbachev}, recasting it in the language of density operators and quantum operations.  We then study the consequences of replacing the three-qubit GHZ state in their scheme with a three-qubit W state.  Finally, we investigate how the presence of white noise affects the entanglement teleportation capability of the three-qubit GHZ and W states.  We conclude with some remarks on future research.

We begin with a review of the entanglement teleportation protocol $P_0$ of V. N. Gorbachev and A. I. Trubilko \cite{Gorbachev}.  It involves a sender, Alice, and two receivers, Bob and Cindy.  Alice is in possession of an entangled two-qubit input system 45, and another qubit 3 entangled with both a fourth qubit 1 in Bob's possession, and a fifth qubit 2 in Cindy's possession (i.e. a three-qubit entangled state, $\chi_{123}$).  Initially the composite system 12345 is prepared in a state with density operator, $\sigma^{total}_{12345} = \chi_{123} \otimes \pi_{45}$, where \cite{Explanation1}
\begin{equation}
\pi_{45} = |\psi\rangle_{45}\langle\psi|,\
|\psi\rangle_{45}
= \cos\frac{\theta}{2}|01\rangle_{45} + e^{i\phi}\sin\frac{\theta}{2}|10\rangle_{45},
\end{equation}
$0 < \theta < \pi,\ 0 \leq \phi \leq 2\pi$ are the polar and azimuthal angles respectively.  Here, we use $|0\rangle$ and $|1\rangle$ to denote an orthonormal set of basis states for each qubit.  To teleport the input state $\pi_{45}$ to the joint target system 12 of Bob and Cindy, Alice performs a joint Bell basis measurement on systems 3 and 4, described by operators $I_{12} \otimes \Pi^j_{34} \otimes I_5$, $I_{12}$ and $I_5$ are the identity operators on the composite subsystem 12 and subsystem 5 respectively, $j$ labels the outcome of the measurement,
\begin{equation}
\Pi^1_{34} = |\Phi^+\rangle_{34}\langle\Phi^+|,\
\Pi^2_{34} = |\Phi^-\rangle_{34}\langle\Phi^-|,\
\Pi^3_{34} = |\Psi^+\rangle_{34}\langle\Psi^+|,\
\Pi^4_{34} = |\Psi^-\rangle_{34}\langle\Psi^-|,
\end{equation}
where
$$
|\Phi^{\pm}\rangle_{34} = \frac{1}{\sqrt{2}}(|00\rangle_{34} \pm |11\rangle_{34}),
$$
$$
|\Psi^{\pm}\rangle_{34} = \frac{1}{\sqrt{2}}(|01\rangle_{34} \pm |10\rangle_{34})
$$
are the Bell states.  Next, Alice performs a von Neumann measurement on system 5, described by operators $I_{1234} \otimes \Pi^k_5$, $I_{1234}$ is the identity operator on composite subsystem 1234, $k$ labels the outcome of the measurement,
$$
\Pi^1_5 = |\nu^+\rangle_5\langle\nu^+|,\ 
\Pi^2_5 = |\nu^-\rangle_5\langle\nu^-|,
$$
$$
|\nu^+\rangle_5 
= \cos\frac{\mu}{2}|0\rangle_5 + e^{i\lambda}\sin\frac{\mu}{2}|1\rangle_5,
$$
\begin{equation}
|\nu^-\rangle_5 
= -\sin\frac{\mu}{2}|0\rangle_5 + e^{-i\lambda}\cos\frac{\mu}{2}|1\rangle_5,
\end{equation}
$0 \leq \lambda \leq 2\pi$, $0 \leq \mu \leq \pi$.  If Alice's measurements have outcome $j$ and $k$, she broadcasts her measurement results (three-bit) to Bob and Cindy via a classical channel.  The joint state of Bob and Cindy's target system 12 conditioned on Alice's measurement results is given by
\begin{equation}
\rho^{jk}_{12} = \frac{1}{p_{jk}}{\rm tr}_{345}[(I_{12} \otimes \Pi^j_{34} \otimes \Pi^k_5)(\chi_{123} \otimes \pi_{45})],
\end{equation}
where
\begin{equation}
p_{jk} = {\rm tr}_{12345}[(I_{12} \otimes \Pi^j_{34} \otimes \Pi^k_5)(\chi_{123} \otimes \pi_{45})].
\end{equation}
For Bob and Cindy to successfully complete the teleportation protocol, they perform a classically coordinated $j$- and $k$-dependent unitary operation $U^{jk}_{12}$ on system 12 (see Table I and II) such that
\begin{equation}
\tau^{jk}_{12} = U^{jk}_{12}\rho^{jk}_{12}U^{jk\dagger}_{12},
\end{equation}
where $U^{jk}_{12} = \sigma^b_1 \otimes \sigma^c_2$, $b, c = 0, 1, 2, 3$:
$$
\sigma^0 = \left(\begin{array}{cc}
1 & 0 \\ 0 & 1
\end{array}\right),\ \sigma^1 = \left(\begin{array}{cc}
0 & 1 \\ 1 & 0
\end{array}\right),\ \sigma^2 = \left(\begin{array}{cc}
0 & -i \\ i & 0
\end{array}\right),\ \sigma^3 = \left(\begin{array}{cc}
1 & 0 \\ 0 & -1
\end{array}\right).
$$
The success of the entanglement teleportation scheme can be measured by the amount of entanglement transmitted, and the fidelity \cite{Jozsa} between the input state $\pi_{in}$ and the output state $\tau^{jk}_{out}$, averaged over all possible Alice's measurement outcomes, $j$ and $k$, and over an isotropic distribution of input states $\pi_{in}$, given in Eq.(1):
\begin{equation}
\langle F\rangle = 
\frac{1}{4\pi}\int^{\pi}_0\int^{2\pi}_0\sin\theta d\theta d\phi\
\sum^4_{j = 1}\sum^2_{k = 1}p_{jk}F^{jk}
\end{equation}
where
\begin{equation}
F^{jk} \equiv {\rm tr}(\tau^{jk}_{out}\pi_{in}).
\end{equation}
To quantify the amount of entanglement associated with a density matrix $\rho_{SS'}$, we consider the concurrence \cite{Wootters, Hill},
$C = \max\{ \lambda_1 - \lambda_2 - \lambda_3 - \lambda_4, 0\}$ where
$\lambda_k (k = 1,2,3,4)$
are the square roots of the eigenvalues in decreasing order of magnitude of the spin-flipped density matrix operator
$R = \rho_{SS'} (\sigma^2 \otimes \sigma^2) \rho^{\ast}_{SS'} (\sigma^2 \otimes \sigma^2)$, where the asterisk indicates complex conjugation.  For instance, after some straightforward algebra, the concurrence associated with the input state $\pi_{45}$ is given by
$$
C(\pi_{45}) = \max\{\sin\theta,\ 0\}.
$$

In Ref.\cite{Gorbachev}, V. N. Gorbachev and A. I. Trubilko considered
\begin{equation}
\chi_{123} = \chi^{GHZ}_{123} = |GHZ\rangle_{123}\langle GHZ|,\
|GHZ\rangle_{123} = \frac{1}{\sqrt{2}}(|000\rangle_{123} + |111\rangle_{123}).
\end{equation}
Substituting Eq.(9) into Eq.(5) yields
$$
p_{j1} = \frac{1}{8}(1 - \cos\mu\cos\theta),
$$
\begin{equation}
p_{j2} = \frac{1}{8}(1 + \cos\mu\cos\theta).
\end{equation}
It follows from Eq.(1) and results from Eq.(6) that
$$
F^{j1} = \frac{3 - 4\cos\mu\cos\theta + \cos2\theta + 2\cos\lambda\sin\mu\sin^2\theta}{4(1 - \cos\mu\cos\theta)},
$$
\begin{equation}
F^{j2} = \frac{3 + 4\cos\mu\cos\theta + \cos2\theta + 2\cos\lambda\sin\mu\sin^2\theta}{4(1 + \cos\mu\cos\theta)}.
\end{equation}
Putting Eq.(10) and Eq.(11) into Eq.(7) gives
\begin{equation}
\langle F\rangle = \frac{2}{3} + \frac{1}{3}\cos\lambda\sin\mu.
\end{equation}
The average teleportation fidelity $\langle F\rangle$ is thus dependent on Alice's von Neumann measurement on system 5, specified by $\lambda$ and $\mu$.  When $\lambda = 0$, $\mu = \frac{\pi}{2}$, $F^{jk} = 1$ for all $j$ and $k$, and we have $\langle F\rangle = 1$.  The concurrences associated with the output states $\tau^{j1}_{out}$ and $\tau^{j2}_{out}$ are respectively given by
$$
C(\tau^{j1}_{out}) = \max\left\{
\frac{\sin\mu\sin\theta}{1 - \cos\mu\cos\theta},\ 0\right\},
$$
\begin{equation}
C(\tau^{j2}_{out}) = \max\left\{
\frac{\sin\mu\sin\theta}{1 + \cos\mu\cos\theta},\ 0\right\}.
\end{equation}
We note that for each $0 < \theta < \pi$, $C(\tau^{jk}_{out}) = \sin\theta$ is not maximal when $\mu = \frac{\pi}{2}$.  In fact, $C(\tau^{j1}_{out}) = 1$ when $\mu = \theta$, or $C(\tau^{j2}_{out}) = 1$ when $\mu = \pi - \theta$.  However, unless $\theta$ is known, these maxima cannot be obtained.

Here, we consider
\begin{equation}
\chi_{123} = \chi^W_{123} = |W\rangle_{123}\langle W|,\
|W\rangle_{123} = \frac{1}{\sqrt{3}}(|001\rangle_{123} + |010\rangle_{123} + |100\rangle_{123}).
\end{equation}
Substituting Eq.(14) into Eq.(5) yields
$$
p_{11} = p_{21} = \frac{1}{24}[3 + \cos\theta - \cos\mu(1 + 3\cos\theta)],
$$
$$
p_{31} = p_{41} = \frac{1}{24}[3 - \cos\theta + \cos\mu(1 - 3\cos\theta)],
$$
$$
p_{12} = p_{22} = \frac{1}{24}[3 + \cos\theta + \cos\mu(1 + 3\cos\theta)],
$$
\begin{equation}
p_{32} = p_{42} = \frac{1}{24}[3 - \cos\theta - \cos\mu(1 - 3\cos\theta)].
\end{equation}
It follows from Eq.(1) and results from Eq.(6) that
$$
F^{11} = F^{21} = \frac{4\cos^2\frac{\mu}{2}\sin^4\frac{\theta}{2}}{3 + \cos\theta - \cos\mu(1 + 3\cos\theta)},
$$
$$
F^{31} = F^{41} = \frac{4\cos^2\frac{\mu}{2}\sin^2\frac{\theta}{2}(1 + \cos\phi\sin\theta)}{3 - \cos\theta + \cos\mu(1 - 3\cos\theta)},
$$
$$
F^{12} = F^{22} = \frac{4\cos^2\frac{\mu}{2}\cos^2\frac{\theta}{2}(1 + \cos\phi\sin\theta)}{3 + \cos\theta + \cos\mu(1 + 3\cos\theta)},
$$
\begin{equation}
F^{32} = F^{42} = \frac{4\cos^2\frac{\mu}{2}\cos^4\frac{\theta}{2}}{3 - \cos\theta - \cos\mu(1 - 3\cos\theta)}.
\end{equation}
Putting Eq.(15) and Eq.(16) into Eq.(7) gives
\begin{equation}
\langle F\rangle = \frac{5}{9}\cos^2\frac{\mu}{2}.
\end{equation}
The average teleportation fidelity $\langle F\rangle$ is therefore, in contrast to Eq.(12), independent of $\lambda$.  When $\mu = 0$, $\langle F\rangle = \frac{5}{9} < \frac{2}{3}$, which is not better than any classical communication protocol \cite{Bennett, Popescu, Horodecki}.  However, we note from the following
$$
C(\tau^{11}_{out}) = C(\tau^{21}_{out}) = \max\left\{
\frac{8\sin^2\frac{\mu}{2}\cos^2\frac{\theta}{2}}{3 + \cos\theta - \cos\mu(1 + 3\cos\theta)},\ 0\right\},
$$
$$
C(\tau^{31}_{out}) = C(\tau^{41}_{out}) = \max\left\{
\frac{8\cos^2\frac{\mu}{2}\sin^2\frac{\theta}{2}}{3 - \cos\theta + \cos\mu(1 - 3\cos\theta)},\ 0\right\},
$$
$$
C(\tau^{12}_{out}) = C(\tau^{22}_{out}) = \max\left\{
\frac{8\cos^2\frac{\mu}{2}\cos^2\frac{\theta}{2}}{3 + \cos\theta + \cos\mu(1 + 3 \cos\theta)},\ 0\right\},
$$
\begin{equation}
C(\tau^{32}_{out}) = C(\tau^{42}_{out}) = \max\left\{
\frac{8\sin^2\frac{\mu}{2}\sin^2\frac{\theta}{2}}{3 - \cos\theta - \cos\mu(1 - 3\cos\theta)},\ 0\right\},
\end{equation}
that with total probability
$$
p_{31} + p_{41} + p_{12} + p_{22} = \frac{1}{6}(3 + \cos\mu) = \frac{2}{3},
$$
independent of $0 < \theta < \pi$, we have
$$
C(\tau^{31}_{out}) = C(\tau^{41}_{out}) = C(\tau^{12}_{out}) = C(\tau^{22}_{out}) = 1.
$$
In fact, it can be shown that $\tau^{31}_{out} = \tau^{41}_{out} = \tau^{12}_{out} = \tau^{22}_{out} = |\Psi^+\rangle$, that is, a Bell state.  So, as long as the input state $\pi_{in}$ has nonzero entanglement \cite{Explanation1}, we could end up having $|\Psi^+\rangle$ as output state, with total probability equal to $\frac{2}{3}$, when $\mu = 0$.  In view of this, it would be more appropriate to call this process ``probabilistic entanglement swapping'' \cite{Ekert2, Zeilinger2}.

Finally, we study the effects on the Gorbachev-Trubilko entanglement teleportation protocol $P_0$ \cite{Gorbachev} due to the presence of white noise.  In particular, we want to see the effect on the average teleportation fidelity and amount of entanglement transmitted when the three-qubit GHZ state is used as a channel.  For the three-qubit W channel, we consider, for $\mu = 0$, the effect on the total probability of having an entangled output state $\pi^{jk}_{out}$ and its associated concurrence.  To these ends, we consider
\begin{equation}
\hat{\chi}^{GHZ}_{123} = w\chi^{GHZ}_{123} + (1 - w)\frac{1}{8}I_{123},
\end{equation}
and
\begin{equation}
\hat{\chi}^W_{123} = w\chi^W_{123} + (1 - w)\frac{1}{8}I_{123},
\end{equation}
where $0 \leq w \leq 1$ is called visibility in Ref.\cite{Dagomir}.  It defines to what extent the quantum processes associated with $\chi^{GHZ}_{123}$ $(\chi^W_{123})$ are visible in those given by $\hat{\chi}^{GHZ}_{123}$ $(\hat{\chi}^W_{123})$.  If $w = 0$, no trace is left of these $\chi^{GHZ}_{123}$ $(\chi^W_{123})$ generated processes, and if $w = 1$, we have the full visibility, not affected by any noise.  Substituting $\chi_{123}$ in Eq.(4) with Eq.(19) and going through essentially the same calculations, we obtain for the three-particle GHZ state,
\begin{equation}
\langle F\rangle = \left(\frac{2}{3} + \frac{1}{3}\cos\lambda\sin\mu\right) - \left(\frac{5}{12} + \frac{1}{3}\cos\lambda\sin\mu\right)(1 - w),
\end{equation}
$$
C(\tau^{j1}_{out}) = \max\left\{
\frac{\sin\mu\sin\theta}{1 - \cos\mu\cos\theta} - \left(\frac{1}{2} + \frac{\sin\mu\sin\theta}{1 - \cos\mu\cos\theta}\right)(1 - w),\
0\right\},
$$
\begin{equation}
C(\tau^{j2}_{out}) = \max\left\{
\frac{\sin\mu\sin\theta}{1 + \cos\mu\cos\theta} - \left(\frac{1}{2} + \frac{\sin\mu\sin\theta}{1 + \cos\mu\cos\theta}\right)(1 - w),\
0\right\}.
\end{equation}
The rate of decrease of $\langle F\rangle$ $(C(\tau^{jk}_{out}))$ with respect to $(1 - w)$ depends on Alice's single-qubit measurement on system 5, specified by $\lambda$ and $\mu$.  That is, the white noise couples to Alice's von Neumann measurement, Eq.(3).  $\langle F\rangle$ $(C(\tau^{jk}_{out}))$ decreases at a higher rate when Alice's measurement yields a higher $\langle F\rangle$ $(C(\tau^{jk}_{out}))$.  For $\mu = \frac{\pi}{2}$ and $\theta = \frac{\pi}{2}$, it follows from Eq.(22),
\begin{equation}
C(\tau^{j1}_{out}) = C(\tau^{j2}_{out}) = \max\left\{
1 - \frac{3}{2}(1 - w),\ 0\right\}.
\end{equation}
Putting Eq.(20) into Eq.(4), we have for the three-particle W state, when $\mu = 0$,
\begin{equation}
C(\tau^{31}_{out}) = C(\tau^{41}_{out}) = C(\tau^{12}_{out}) = C(\tau^{22}_{out}) = \max\left\{1 - \frac{9(1 - w)}{2[4 - (1 - w)]},\ 0\right\},
\end{equation}
with total probability
\begin{equation}
p_{31} + p_{41} + p_{12} + p_{22} = \frac{2}{3} - \frac{1}{6}(1 - w).
\end{equation}
So, although the total probability decreases linearly with increasing $(1 - w)$, it is never zero.  The concurrence associated with the entangled output states decreases at an increasing rate with respect to $(1 - w)$ and is nonzero as long as $(1 - w) < \frac{8}{11}$, in contrast to Eq.(23).

In conclusion, we recast the entanglement teleportation scheme $P_0$ of V. N. Gorbachev and A. I. Trubilko \cite{Gorbachev} in the language of density operators and quantum operations.  This allows us to investigate the consequences of replacing the three-qubit GHZ state in their original scheme with a three-qubit W state, both in terms of the average teleportation fidelity and the amount of entanglement transmitted.  Our results show that, $P_0$ with a three-qubit W channel, is impossible to give average fidelity better than any classical communication protocol.  It is not clear if there exist more general completely positive maps rather than the unitary operators $U^{jk}_{12}$, which could yield a higher average teleportation fidelity.  However, we show that with total probability $\frac{2}{3}$, we could have a maximally entangled Bell state as output, if the input state has nonzero entanglement.  We are also able to study the effects of white noise.  It would be interesting to see how the average fidelities and amount of entanglement transmitted would change when the GHZ and W states are being subjected to other types of noise.

The author thanks A. I. Trubilko for useful pointers, Yuri Suhov and Andrew Skeen for useful discussions.  This publication is an output from project activity funded by The Cambridge MIT Institute Limited (``CMI'').  CMI is funded in part by the United Kingdom Government.  The activity was carried out for CMI by the University of Cambridge and Massachusetts Institute of Technology.  CMI can accept no responsibility for any information provided or views expressed.

\newpage

\begin{table}
\begin{ruledtabular}
\begin{tabular}{ccc}
Alice's measurement results $j$, $k$ & Bob and Cindy's unitary operation $U^{jk}_{12}$\\
1, 1 & $\sigma^0_1 \otimes \sigma^1_2$ or $\sigma^3_1 \otimes \sigma^2_2$ \\
2, 1 & $\sigma^0_1 \otimes \sigma^2_2$ or $\sigma^3_1 \otimes \sigma^1_2$ \\
3, 1 & $\sigma^1_1 \otimes \sigma^0_2$ or $\sigma^2_1 \otimes \sigma^3_2$ \\
4, 1 & $\sigma^2_1 \otimes \sigma^0_2$ or $\sigma^1_1 \otimes \sigma^3_2$ \\
1, 2 & $\sigma^0_1 \otimes \sigma^2_2$ or $\sigma^3_1 \otimes \sigma^1_2$ \\
2, 2 & $\sigma^0_1 \otimes \sigma^1_2$ or $\sigma^3_1 \otimes \sigma^2_2$ \\
3, 2 & $\sigma^2_1 \otimes \sigma^0_2$ or $\sigma^1_1 \otimes \sigma^3_2$ \\
4, 2 & $\sigma^1_1 \otimes \sigma^0_2$ or $\sigma^2_1 \otimes \sigma^3_2$
\end{tabular}
\end{ruledtabular}
\caption{\label{I}}
Bob and Cindy's unitary operations conditioned on Alice's measurement results, when Alice, Bob and Cindy share a three-qubit entangled GHZ state.
\end{table}

\begin{table}
\begin{ruledtabular}
\begin{tabular}{ccc}
Alice's measurement results $j$, $k$ & Bob and Cindy's unitary operation $U^{jk}_{12}$\\
1, 1 & $\sigma^1_1 \otimes \sigma^0_2$ or $\sigma^1_1 \otimes \sigma^3_2$ or $\sigma^2_1 \otimes \sigma^0_2$ or $\sigma^2_1 \otimes \sigma^3_2$ \\
2, 1 & $\sigma^1_1 \otimes \sigma^0_2$ or $\sigma^1_1 \otimes \sigma^3_2$ or $\sigma^2_1 \otimes \sigma^0_2$ or $\sigma^2_1 \otimes \sigma^3_2$ \\
3, 1 & $\sigma^0_1 \otimes \sigma^0_2$ or $\sigma^1_1 \otimes \sigma^1_2$ or $\sigma^2_1 \otimes \sigma^2_2$ or $\sigma^3_1 \otimes \sigma^3_2$ \\
4, 1 & $\sigma^0_1 \otimes \sigma^0_2$ or $\sigma^1_1 \otimes \sigma^1_2$ or $\sigma^2_1 \otimes \sigma^2_2$ or $\sigma^3_1 \otimes \sigma^3_2$ \\
1, 2 & $\sigma^0_1 \otimes \sigma^0_2$ or $\sigma^1_1 \otimes \sigma^1_2$ or $\sigma^2_1 \otimes \sigma^2_2$ or $\sigma^3_1 \otimes \sigma^3_2$ \\
2, 2 & $\sigma^0_1 \otimes \sigma^0_2$ or $\sigma^1_1 \otimes \sigma^1_2$ or $\sigma^2_1 \otimes \sigma^2_2$ or $\sigma^3_1 \otimes \sigma^3_2$ \\
3, 2 & $\sigma^0_1 \otimes \sigma^1_2$ or $\sigma^0_1 \otimes \sigma^2_2$ or $\sigma^3_1 \otimes \sigma^1_2$ or $\sigma^3_1 \otimes \sigma^2_2$ \\
4, 2 & $\sigma^0_1 \otimes \sigma^1_2$ or $\sigma^0_1 \otimes \sigma^2_2$ or $\sigma^3_1 \otimes \sigma^1_2$ or $\sigma^3_1 \otimes \sigma^2_2$
\end{tabular}
\end{ruledtabular}
\caption{\label{II}}
Bob and Cindy's unitary operations conditioned on Alice's measurement results, when Alice, Bob and Cindy share a three-qubit entangled W state.
\end{table}


\begin{thebibliography}{99}
\bibitem{Explanation1} In Ref.\cite{Weber}, L. Marinatto and T. Weber have shown that a two-qubit pure state of general type cannot be perfectly teleported through a three-qubit quantum channel.  In particular, $P_0$ does not work if $\pi_{45}$ has zero entanglement.  In this paper, for simplicity, we consider the same class of input states as in Ref.\cite{Gorbachev}.
\end{thebibliography}
\end{document}